\documentclass{naturedummy}
\usepackage{amsmath}
\usepackage{pdfsync}
\usepackage{graphicx}

\advance \topmargin by -\headheight
\advance \topmargin by -\headsep

\evensidemargin \oddsidemargin
\newcommand{\ket}[1]{\left\vert#1\right\rangle}
\newcommand{\bra}[1]{\left\langle#1\right\vert}

\newcommand{\up}{\left\vert \uparrow\right\rangle}
\newcommand{\down}{\left\vert \downarrow\right\rangle}
\newcommand{\expectation}[1]{\langle #1 \rangle }

\newcommand{\kupup}{\ket{\uparrow\uparrow}}
\newcommand{\kupdown}{\ket{\uparrow\downarrow}}
\newcommand{\kdownup}{\ket{\downarrow\uparrow}}
\newcommand{\kdowndown}{\ket{\downarrow\downarrow}}
\newcommand{\bupdown}{\bra{\uparrow\downarrow}}
\newcommand{\bdownup}{\bra{\downarrow\uparrow}}
\newcommand{\upup}{\uparrow\uparrow}
\newcommand{\updown}{\uparrow\downarrow}
\newcommand{\downup}{\downarrow\uparrow}
\newcommand{\downdown}{\downarrow\downarrow}

\newcommand{\Swap}{\textrm{swap}}
\newcommand{\Parity}{\textrm{parity}}
\newcommand{\Visibility}{\textrm{visibility}}

\newcommand{\Sr}{$^{88}$Sr$^+$\ }

\title{Measurement of the magnetic interaction between two electrons}

\author{Shlomi Kotler$^{1}$\footnote{Current address: Physical Measurement Laboratory, National Institute of Science and Technology, Boulder CO, 80305, USA.}, Nitzan Akerman$^1$, Nir Navon$^1$\footnote{Current address: Cavendish Laboratory, University of Cambridge, J.J. Thomson Avenue, Cambridge CB30HE, United Kingdom.}, Yinnon Glickman$^1$, \& Roee Ozeri$^1$}

\begin{document}

\maketitle

\begin{affiliations}
 \item Department of Physics of Complex Systems, Weizmann Institute of Science, Rehovot 76100, Israel.
\end{affiliations}

\begin{abstract}
Electrons have an intrinsic, indivisible, magnetic dipole aligned with their internal angular momentum (spin)\cite{UhlenbeckGoudsmit1926}. The magnetic interaction between two electrons can therefore impose a change in their spin orientation. Similar dipolar magnetic interactions exists between other spin systems and were studied experimentally. Examples include the interaction between an electron and its nucleus or between several multi-electron spin complexes\cite{BudkerDeMill2004,Mills1975,Ritter1984,Pfau2005,Lev2011,Wrachtrup2013,Yacoby2013}. The process for two electrons, however, was never observed in experiment. The challenge is two-fold. At the atomic scale, where the coupling is relatively large, the magnetic interaction is often overshadowed by the much larger coulomb exchange counterpart\cite{BudkerDeMill2004}. In typical situations where exchange is negligible, magnetic interactions are also very weak and well below ambient magnetic noise. Here we report on the first measurement of the magnetic interaction between two electronic spins. To this end, we used the ground state valence electrons of two \Sr ions, co-trapped in an electric Paul trap and separated by more than two micrometers. We measured the weak, millihertz scale (alternatively $10^{-18}$ eV or $10^{-14}$ K), magnetic interaction between their electronic spins. This, in the presence of magnetic noise that was six orders of magnitude larger than the respective magnetic fields the electrons apply on each other. Cooperative spin dynamics was kept coherent for $15\ $s during which spin-entanglement was generated, verified by a negative measured value of $-0.16(2)$ for the swap entanglement witness. The sensitivity necessary for this measurement was provided by restricting the spin evolution to a Decoherence-Free Subspace (DFS) which is immune to collective magnetic field noise. Finally, by varying the separation between the two ions, we were able to recover the inverse cubic distance dependence of the interaction. The reported method suggests an alternative route to the search of long-range anomalous spin-spin forces\cite{MoodyWilczek1984} and can be generalized to include Quantum Error Correction codes\cite{Retzker2013,Ozeri2013,Kraus2013,Ye2013} for other cases of extremely weak signal detection.
\end{abstract}

Early during the twentieth century, a number of experiments indicated that the electron is more than just an electrically charged point-particle. By introducing the electron spin and its accompanying magnetic moment, Goudsmit and Uhlenbeck\cite{UhlenbeckGoudsmit1926} explained a multitude of experimental observations such as the fine-structure spectrum of hydrogen, the anomalous Zeeman splitting, as well as the famous Stern-Gerlach experiment. Since then, the magnetic field of a single electron was detected\cite{Rugar2004} and its magnetic dipole measured with unprecedented accuracy\cite{Gabrielse2008}.

Since a single electron is a tiny magnet, every two electrons should influence each others magnetic dipole orientation, as magnets do. Although fundamental in nature, a direct measurement of this magnetic dipolar interaction was never performed. Evidence for electron spin-spin interaction is abundant at the atomic scale. These, however, do not result from magnetic torques but rather involve the electron charge. Spin correlations in Helium-like atoms, in the formation of solid ferromagnets as well as in covalent chemical bonds, among other examples, are all accounted for by Coulomb exchange interaction\cite{BudkerDeMill2004}. They originates from the fact that electrons are indistinguishable charged fermions. Whenever the charge densities of two electrons overlap significantly, Coulomb spin-exchange energy overwhelms the magnetic counterpart. Particles do exhibit magnetic spin-spin interaction at the atomic scale as long as they are not identical, so exchange plays no role. Examples include the electron-proton magnetic interaction, responsible for the well known and extensively measured hyperfine splitting of atomic energy levels\cite{BudkerDeMill2004}, or the magnetic interaction between an electron and an anti-electron inside Positronium\cite{Mills1975,Ritter1984}.

The problem of observing magnetic interaction on top of exchange forces is common to other systems where identical spins are involved. It can be resolved by increasing the inter-spin separation $d$. Although the magnetic energy becomes dominant, it also decreases with distance, scaling as $d^{-3}$. Therefore, such an approach can only be fruitful when accompanied by an appropriate increase in the magnetic dipole moment or an improvement in the measurement sensitivity. With recent advances in magnetometry at the tens of nanometer scale, the magnetic interaction of two Nitrogen Vacancy (NV) spin-$1$ defects in diamonds has been observed to result in their entanglement\cite{Wrachtrup2013}, measuring weak interaction strengths\cite{Yacoby2013}, as low as $60$ Hz. A comparable magnetic interaction strength was observed between atoms in dipolar quantum gases\cite{Pfau2005,Lev2011}. In these cold gases, the relatively large inter atomic distance of hundreds of nanometers was compensated by the large magnetic dipole of each atom, ranging from six to ten times that of the electron.

We measured the weak magnetic dipolar coupling between two electrons at the micrometer separation scale, using atomic ions. Here we used two trapped \Sr ions, each having a single valence electron and no nuclear spin. These bound electrons inherited the well isolated environment of their ions along with a high degree of controllability. Indeed, ions can be tightly confined and laser-cooled to their mechanical ground state\cite{Bible1998}, allowing for the long interrogation times necessary for weak signal measurements. Examples include state-of-the-art detection of electric\cite{Roos2006,Turchette2000,Biercuk2010} and magnetic\cite{Roos2004,Langer2005,Kotler2011,FSK2012,Kotler2013DD} fields, as well as gravity\cite{Chou2010}. The relative magnetic energy correction imposed by using bound rather than free electrons\cite{Breit1928}, is smaller than $0.004\%$ and well below our reported sensitivity.

Essentially, our apparatus enabled us to place the electronic spins at a controlled distance from one another, as well as initialize, manipulate and detect their internal spin state with high fidelity. Details of the setup are found in Ref \cite{Akerman2011} as well as in the Supplementary Information. Briefly, a Coulomb crystal of two ions was formed in an electrical Paul trap\cite{Bible1998}. We used external voltages to push the ions against their Coulomb repulsion (see Figure 1a), thus controlling the inter-ion separation $d$. The minimal distance attained was limited by our ability to maintain stable ion-crystals without incurring a trap voltage breakdown. The inter-ion distance $d$ is the difference between the equilibrium positions of two charged particles trapped in a harmonic trap, $d=(2k_ee^2/M(2\pi f_{trap})^2)^{1/3}$, where $k_e$ is Coulomb's constant, $e$ the electron charge and $M$ is the mass of \Sr. The oscillation frequency $f_{trap}$ was measured spectroscopically. For \Sr, the valence electron spin states are $\up=\ket{5s_{1/2}, J=1/2, M_J=1/2}$ and $\down=\ket{5s_{1/2}, J=1/2, M_J=-1/2}$. State initialization to $\kupup$ was done by optical pumping. We were able to perform all possible collective spin rotations by pulsing a resonant radio frequency (rf) magnetic field and tuning the pulse duration and the rf field phase. State detection was performed by state-selective fluorescence, distinguishing $\kupup$ from $\kdowndown$ from either $\kdownup$ or $\kupdown$, the latter two were indistinguishable\cite{Keselman2011}. All these collective operations had better than $98\%$ typical fidelities. We utilized inhomogeneities in the ion trap potential to perform differential spin rotations\cite{Ulrich2013,Navon2013}, and were able to generate, for example, $\kupdown$ with typically better than $98\%$ fidelity. Finally we were able to generate the entangled states $\ket{\Psi_{\pm}}=(\kupdown\pm\kdownup)/\sqrt{2}$ using a S\o rensen-M\o lmer entangling gate with typically $95\%$ fidelity\cite{Navon2014}.



We now turn to describe the magnetic dipolar interaction and competing noise. As shown in Figure 1a, we aligned the external magnetic field along the line connecting the two ions. In the case of a uniform magnetic field $B$, the spin part of the two-ion Hamiltonian can be written as,
\begin{equation}\label{hamiltonian}
H = \frac{\hbar\omega_A}{2}(\sigma_{z,1} + \sigma_{z,2}) + 2\hbar\xi\sigma_{z,1}\sigma_{z,2} - \hbar\xi(\sigma_{x,1}\sigma_{x,2} + \sigma_{y,1}\sigma_{y,2}).
\end{equation}
Here, $\hbar$ is the Planck constant divided by $2\pi$, $\sigma_{j,i}$ is the $j\in\{x,y,z\}$ Pauli spin operator of the $i$'th spin, $\omega_A = \frac{g\mu_{B}B}{2\hbar}$ is the spin Larmor frequency where $\mu_B$ and $g$ are the Bohr magneton and the electron spin gyromagnetic ratio, respectively. The spin-spin interaction strength is $\xi=\mu_0(g\mu_B/2)^2/4\pi \hbar d^3$ with $\mu_0$ the vacuum permeability constant. The first term on the right-hand-side of Eq. \ref{hamiltonian} describes the Zeeman shift of the spins energy, due to a uniform external magnetic field. The second and third terms are due to spin-spin interactions. The second term creates a shift in the resonance frequency of one spin that is conditioned on the state of the other, and was recently measured for the case of two NV spin-1 defects \cite{Wrachtrup2013,Yacoby2013}. The third term results in a collective spin flip in which a spin excitation is exchanged. Due to conservation of energy, for this term to be on-resonance and effective, the two spins have to be exactly degenerate, i.e. $B$ has to be exactly uniform. It is the third term which was at the focus of our experiment.

Ultimately, the ability to measure a weak magnetic spin-spin interaction is limited by collective external magnetic field fluctuations, described by the first term in Eq. 1. Typical laboratory magnetic field noise amplitude are on the order of $0.1\ \mu$T, causing fluctuations in $\omega_A$ on the order of a few kilohertz. These are, unfortunately, six orders of magnitude greater than the spin-spin interaction strength.

A state-space solution can remedy the effect of these large magnetic fluctuations. It requires identifying a set of quantum states which are, on the one hand, sensitive to the desired signal, and on the other hand invariant under a certain class of noise processes. Previously this approach was used to measure magnetic field gradients\cite{Roos2004,Langer2005,FSK2012} as well as narrow laser linewidths and the electric-quadrupole of atomic levels \cite{Roos2006}. Here, we tailored the states to the magnetic dipolar interaction. The four eigenstates of the Hamiltonian in Eq. \ref{hamiltonian} are $\kupup,\kdowndown$ and the two entangled states $\ket{\Psi_{\pm}}=(\kupdown\pm\kdownup)/\sqrt{2}$. The first two eigenstates are twice as susceptible to magnetic field fluctuations as compared to the single spin states, whereas the energy splitting between the latter two is $4h\xi$ and does not depend on $B$ at all (see Figure 1b for an energy level diagram). By restricting the spin-spin evolution to the DFS spanned by $\ket{\Psi_{\pm}}$, one can observe spin-spin interactions without being sensitive to spatially homogeneous magnetic noise. This was achieved thanks to our ability to address the spins individually so that their state is initialized to either $\kupdown$ or $\kdownup$, i.e within the DFS. 

Spin-spin interaction within the DFS takes a simple form which can be understood in terms of the geometric Bloch sphere representation shown in Figure 1c. In this subspace, Eq. \ref{hamiltonian} rewrites as $H=2\hbar\xi(\kupdown\bdownup+\kdownup\bupdown)$, up to a global constant. The $\ket{\Psi_{\pm}}$ states are invariant under the interaction (Figure 1b). All other states undergo rotation (Figure 1c, solid blue arc) around the direction defined by $\ket{\Psi_{\pm}}$, hereafter referred to as the $\vec{x}$ direction (Figure 1c). Starting from the north pole ($\kupdown$) the system rotates through the fully entangled state $\ket{\chi_+}=(\kupdown+i\kdownup)/\sqrt{2}$ and toward the south pole ($\kdownup$).

Even in the noise protected subspace, spatial inhomogeneity in the external magnetic field can wash the spin-spin signal away. After reducing inhomogeneities by a factor of a $1000$, we observed residual gradients of $3\times 10^{-7}$T$/$m. This was enough to lift the degeneracy between $\kupdown$ and $\kdownup$ by $\Delta\omega_A=(2\pi)20$ mHz, thus detuning the weak, milihertz, spin-spin coupling from resonance, resulting in a Hamiltonian   $H=2\hbar\xi(\kupdown\bdownup+\kdownup\bupdown)+\hbar\Delta\omega_A/2(\kupdown\bupdown-\kdownup\bdownup)$. In geometric terms, starting at the Bloch sphere north pole, the system state is rapidly rotated by the field gradient about the $\vec{z}$ axis (see red arc in Figure 1c). This counteracts the slower $\vec{x}$ revolution imposed by the spin-spin interaction, restricting its effect to a narrow region of $\sim \pi/400$ steradian solid angle near the north pole.

Using a train of spin-echos, we were able to further reduce these excessive magnetic field inhomogeneities by two orders of magnitude to a negligible level. During their magnetic spin-spin evolution, the two dipoles were flipped at a rate of $f_0=2$ Hz. In geometric terms, this corresponds to a train of $180^\circ$ rotations about the $\vec{x}$ axis (see Figure 1c). These collective rotations do not change the relative orientation of the spins, leaving the spin-spin interaction invariant, as seen in Figure 1d, upper middle three spheres. The effect of the gradient, however, is averaged to zero since exchanging $\kupdown$ and $\kdownup$ is equivalent to constantly switching the sign of the magnetic field gradient (see Figure 1d, lower middle three spheres).

We used parity analysis to obtain a physical observable that was first-order sensitive to the interaction strength. The parity observable measures the coherence between $\kupdown$ and $\kdownup$, so its expectation value for pure states of the form $a\kupdown+b\kdownup$ is $\expectation{\Parity}=a^*b+b^*a$. To measure it, we applied the following experimental sequence (see Figure 1d). The system state is initialized to $e^{i\phi_{init}}\kupdown$, and then evolves under spin-spin interaction to $\ket{\psi(T)}=e^{i\phi_{init}}(\cos(2\xi T) \kupdown +i\sin(2\xi T)\kdownup)$. We then applied a controlled magnetic field gradient, adding a superposition phase $\phi_{parity}$ to $\ket{\psi(T)}=e^{i\phi_{init}}(\cos(2\xi T) \kupdown +ie^{i\phi_{parity}}\sin(2\xi T)\kdownup)$. The parity is then estimated by $\expectation{\Parity}_{est}=P_{\upup}+P_{\downdown}-P_1$, where $P_{\upup},P_{\downdown},P_1\equiv P_{\updown}+P_{\downup}$ are the probabilities to find the system in respective states, measured projectively after performing a collective $\pi/2$ spin rotation (see Supplementary Material). In this case, $\expectation{\Parity}=\sin(4\xi T)\sin(\phi_{parity})$. The parity visibility, $\sin(4\xi T)$, is extracted either by scanning $\phi_{parity}$ (Figure 1e) or by setting it to $\pi/2$. Geometrically, parity corresponds to the projection of the Bloch vector on the $\vec{x}$ axis (rightmost sphere in Figure 1d) and its visibility corresponds to the projection of the Bloch vector on the $x-y$ plane (black double arrow in Figure 1c). The value of $\phi_{init}$ was interleaved between $0$ and $\pi$, using a controlled magnetic gradient. This further reduced the already small effect of initialization imperfection on the parity signal, as explained in the Supplementary Information.

Measuring a weak, millihertz scale, interaction requires a long experiment duration of many seconds. In principal, spin-spin interaction will generate the fully entangled state $\ket{\chi_+}$ after more than a minute of coherent quantum evolution (see Figure 1c). In which case, the parity visibility attains the maximal value of one. The duration of our experiment, however, was limited to $15$ seconds or less, so only partial entanglement was generated and the parity visibility was below $0.4$ (see Figure 1c, black double arrow). The main limiting factor was a reduction in spin detection fidelity as a function of time. It resulted from heating of the ions motion by time-varying electric fields\cite{Turchette2000}. Detection is performed using electron shelving followed by state-selective fluorescence, based on a narrow line-width ($<100\ Hz$) laser\cite{Keselman2011}. Ion heating introduces Doppler shifts on the shelving transition, thereby limiting detection fidelity. As a result, the measured parity observable visibilty reduces by a factor of $\alpha=1-4D(1-D)$ where $D$ is the average of the $\up$ and $\down$ detection fidelities. A further, less significant reduction in $\alpha$ by a factor of $>0.98$ is due to imperfect initialization. See Supplementary Information for the derivation of $\alpha$ and discussion.

To estimate the deterioration in detection fidelity as a function of time we initialized the two-ion crystal to $\kupup$, followed by a wait time with no laser-cooling during which spin-flip modulation at $f_0=2$ Hz is imposed, as in the actual experiment. Finally the system state was detected. Figure 2a shows the deterioration in the detection of $\kupup$ ($\kdowndown$) by the blue (red) dots for an inter-ion distance of $d=2.4\ \mu$m. Detection fidelity degraded from better than $0.95$ at $5$ s to as low as $0.88$ at $T=25$ s. Asymmetry in the detection scheme accounts for the better fidelity of $\kupup$ measurement as compared to $\kdowndown$ (see Supplementary Information). Similar detection fidelities are displayed in Figure 2b, here as a function of ion separation, for a fixed $T=15$ s experiment time.

Even when restricting the experiment duration to $T\le 15$ s, the experiment is long enough for dephasing to potentially limit the observation of spin-spin interaction. Here, dephasing within the DFS, for example due to residual noise in the magnetic field gradient, averages away the superposition relative phase between $\kupdown$ and $\kdownup$. It is represented geometrically by uncontrolled $\vec{z}$ rotations, resulting in an exponential decrease in the parity visibility as a function of the experiment time. To characterize this phase coherence, the system was initialized to $\ket{\Psi_+}=(\kupdown+\kdownup)/\sqrt{2}$ using a M\o lmer-S\o renson entangling gate\cite{Navon2014}. This was followed by a wait of duration $T$ while performing spin flips as in the actual spin-spin experiment, and ended with parity analysis.  The state $\ket{\Psi_+}$ was chosen since it is invariant under spin-spin coupling, while being sensitive to dephasing. Figure 2c, displays the results for $T=0.1\ $s ($T=15\ $s) by the blue (red) circles. A best fit to a cosine yields a parity amplitude of $0.81(5)$ ($0.59(4)$). A conservative estimate for coherence time, not taking detection fidelity into account, yields $44\pm 12 s$. Taking into account the degradation in detection, we cannot observe any statistically-significant dephasing after 15 seconds.

We now turn to describe the main results of this letter.
Figure 3 presents the parity measurements for two electronic spins undergoing magnetic dipolar interaction, at an inter-ion distance of $d=2.4\ \mu$m. A parity oscillation of $\expectation{\Parity}=\pm\alpha\sin(4\xi T)\sin(\phi_{parity})$ is expected, positive when the initial state is $\kupdown$ and negative for the $\kdownup$ initial state. Here, the contrast degradation factor $\alpha=0.68$, is calculated from the data in Figure 2a. Figure 3a shows parity vs. $\phi_{parity}$, for a time $T=0.1$s, which is much shorter than the spin-spin coupling time-scale. As expected, no significant parity oscillation amplitude is detected. The $T=15$ s long experiment results are shown in Figure 3b. Here, the parity sinusoidal dependence becomes evident. The solid blue and red lines are calculated from theory without any adjustable parameters, showing good agreement with the measured data. Shaded areas represent measurement uncertainties in determining $\alpha$. The theoretical interaction strength at the $d=2.4\ \mu$m distance is $\xi=0.93$ mHz, in agreement with a single parameter best fit of the data to the above theory, yielding $\xi=0.9(1)$ mHz. With the parity analysis sinusoidal dependance on $\phi_{parity}$ established, the parity visibility can be measured by fixing $\phi_{parity}=\pi/2$, acquiring a single point rather than a complete sinusoidal fringe. In Figure 3c we display the visibility vs. interaction time (blue circles), in agreement with $\Visibility=\alpha\sin(4\xi T)$, plotted in solid blue.

Although only partial entanglement is generated by spin-spin interaction after $15$ s, it can still be observed by choosing a proper entanglement witness\cite{Horodecki1996}. In such a measurement, a physical observable $S$ is chosen such that its expectation value is positive with respect to all separable (i.e. classical) states. A negative expectation value implies that the system is in an entangled state, reaching $-1$ for a fully entangled state. Here, we chose the swap operator, defined as $\Swap\ket{a,b}=\ket{b,a}$ for any two single spin states $\ket{a}$ and $\ket{b}$. In terms of the two spins density matrix, $\expectation{\Swap}=\rho_{11}+\rho_{44}+\rho_{23}+\rho_{32}=P_{\upup}+P_{\downdown}+\expectation{\Parity}$, where $P_{\upup}$ ($P_{\downdown}$) is the probability of the system to be in the $\kupup$ ($\kdowndown$) state. As $\expectation{\Parity}=\sin(4\xi T)\sin(\phi_{control})$, the minimum of $\expectation{\Swap}$ is, $\expectation{S}=P_{\upup}+P_{\downdown}-\Visibility$. Therefore, entanglement is proven by experimentally verifying the inequality $P_{\upup}+P_{\downdown}<\Visibility$. We repeat the spin-spin experiment $N=2388$ times, at $d=2.4\ \mu$m distance, measuring $\Visibility=0.28(2)$ and $P_{\upup}+P_{\downdown}=0.11(1)$. These conservative estimates, not taking the deterioration in detection fidelity into account, rendered the entanglement witness negative with good statistical significance $\expectation{\Swap}=-0.16(2)$. We assume a projection noise limited error in measured probabilities, supported by an Allen-deviation analysis (ADEV). Taking detection fidelity into account, using the calibration shown in Figure 2a, our maximum likelihood estimate (MLE) renders $\expectation{\Swap}=-0.41(4)$. See Supplementary Information for detailed MLE and ADEV analysis.

Finally, the spin-spin interaction dependence on inter-electron distance is revealed by repeating the above measurement at different ion separations $d$. Figure 4a shows the measured parity visibility (blue circles) vs. $d$. Our measured data is in good agreement with theory (solid blue line): $\Visibility=\alpha\sin(4\xi T)$ where $\alpha$ is calculated from the data shown in Figure 2b. With this theory, the measured parity visibility is then translated into an estimate of the spin-spin coupling constants $\xi$ at different separations, as shown in Figure 4b (blue circles). A best fit to $\xi=\mu_0(g\mu_B/2)^2/4\pi\hbar d^n$ yields $n=3.0(4)$, in agreement with the cubic dependence of magnetic spin-spin interaction.

An accurate measurement of the magnetic interaction between electrons can contribute to the search for anomalous spin forces. Some extensions of the standard model predict deviations of spin-spin interaction due to the introduction of new weakly interacting boson fields\cite{MoodyWilczek1984}. Experiments conducted in a variety of systems, including trapped ions\cite{Wineland1991}, placed lower bounds on the anomalous interaction strength at different length scales. Currently, the best known bound for anomalous spin-spin interaction at the $\mu$m scale can be inferred from hyperfine spectroscopy in Positonium\cite{Mills1975,Ritter1984}. Although there, the interaction involves an electron and a positron separated by an atomic distance, it can nonetheless be used to place a $2.4\times 10^{-4}$ bound on the ratio between anomalous and normal spin-spin interaction, at the micron scale, by assuming a pseudo-scalar mediated interaction. Our statistical significance for the parity observable could be used to set a corresponding bound of $7\times 10^{-2}$. Albeit of larger uncertainty, the measurements described here were for two electrons and were directly performed at the $\mu$m scale. Our method could therefore be used to bound more general anomalous spin interactions at this length scale. This would necessitate a careful study of systematic effects.

In this work we have used a combination of techniques originally developed for the protection of quantum information, in order to measure a very weak interaction. The usage of DFS and spin-echo techniques allowed us to overcome noise processes that were six orders of magnitude larger than the electron-electron magnetic interaction, enabling a $15$ seconds long coherent experiment and the measurement of a milihertz coupling strength. Decoherence free subspaces are a special case of Quantum Error Correction (QEC) codes. Our reported method can be generalized to include the use of QEC codes for the purpose of measuring small signals in the presence of strong noise. Whether or not this is possible depends on the commutation relation of noise and signal operators\cite{Retzker2013,Ozeri2013,Kraus2013,Ye2013}.




		\includegraphics[scale=0.95]{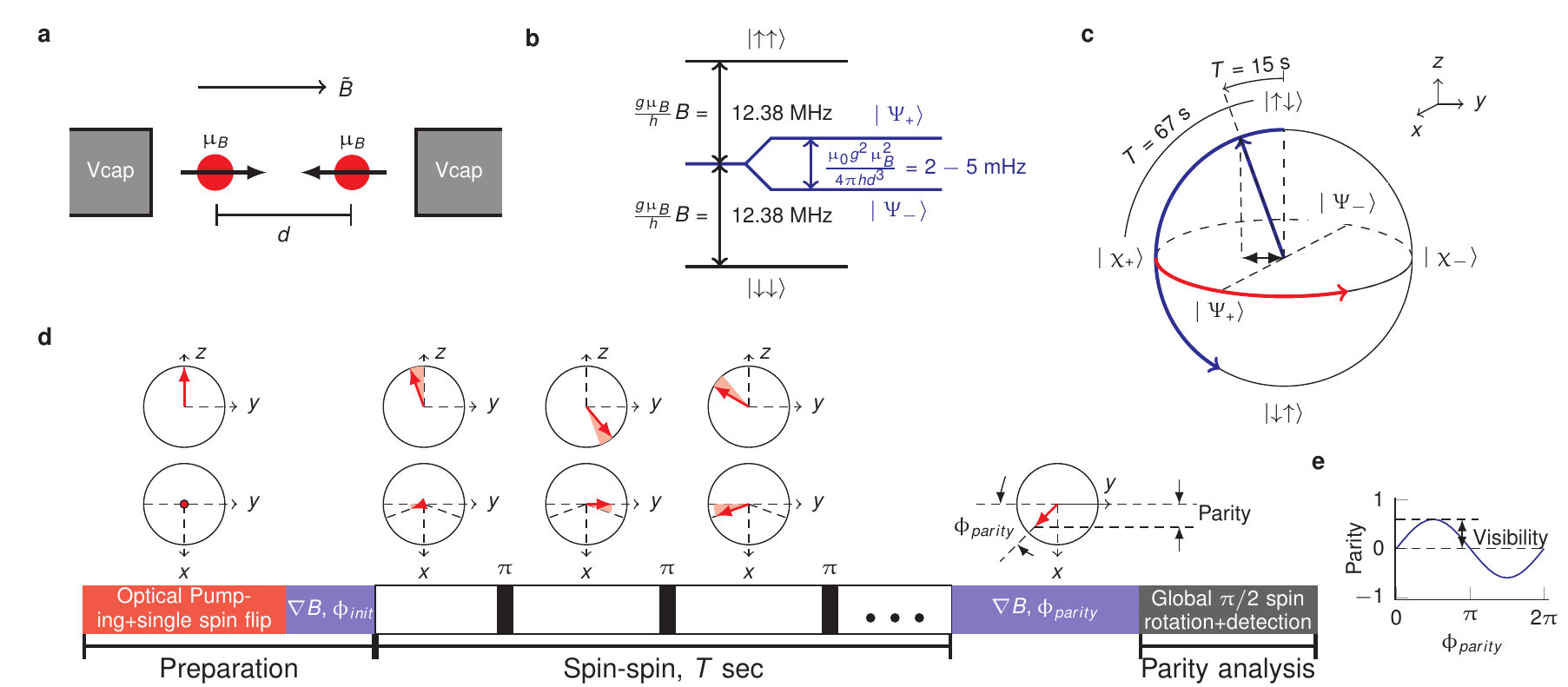}
\begin{figure} 
    \caption{\small
    Experiment overview.
    \textbf{a}, Setup schematics. Two \Sr ions are trapped in a linear rf Paul trap (rf electrodes not shown). The ions are placed at a distance $d$ of $2-3$ microns from one another, controlled by applying $V_{cap}$ voltage to two opposing dc electrodes. The valence electron of each ion has a magnetic moment of nearly one Bohr magneton, $\mu_{B}$. An external magnetic field of $B=0.44\ $mT is aligned along the trap symmetry axis so as to maximize the electron spin-spin interaction.     \textbf{b}, Two spins energy diagram. The external magnetic field splits the states $\up$ (pointing opposite to $B$) and $\down$ (pointing along $B$) by a frequency of $12.34\ MHz$. Magnetic electron spin-spin interaction splits the $\ket{\Psi_{\pm}}=(\kupdown\pm\kdownup)/\sqrt{2}$ states by $4\xi/2\pi=\mu_0(g\mu_B/2)^2/\pi h d^3$, here in the $2-5\ mHz$ range.
    \textbf{c}, Geometric Bloch representation of the DFS subspace spanned by $\kupdown,\kdownup$. The states invariant under spin-spin interaction, $\ket{\Psi_{\pm}}$, are along $\pm\vec{x}$. Correspondingly, the entangled states $\ket{\chi_{\pm}}=(\kupdown\pm i\downup)/\sqrt{2}$ are along $\pm\vec{y}$. Spin-spin interaction induces a counter-clockwise rotation around the $\vec{x}$ axis, shown by the blue arc. Magnetic field gradients generate rotations around the $\vec{z}$ axis, shown by the red arc. For an inter-ion distance of $d=2.4\mu$m (coupling strength of $\xi=(2\pi)0.93$mHz), spin-spin interaction will rotate the state from $\kupdown$ to the fully entangled $\ket{\chi_+}$ state after $T=67\ s$. In our experiments, $T=15\ s$, corresponding to a $21.6^\circ$ rotation. This angle is estimated by the parity visibility (explained below), corresponding to the length of the Bloch vector projection on the $x-y$ plane, indicated by the black double arrow. A collective spin-flip ($\pi$ pulse) corresponds to $180^\circ$ rotation about $\vec{x}$.
    \textbf{d}, Experimental sequence. Corresponding infinitesimal spin evolution is depicted by the shaded red sectors of the $z-y$ and $x-y$ projections of the Bloch sphere. The system is prepared in the $\kupdown$ state by optical pumping to $\kupup$ followed by a single spin flip. Spin-spin evolution is interrupted by short ($9-10\ \mu s$) equidistant collective spin flips ($\pi$ pulses), restricting the effect of magnetic field gradients, as shown by the bottom middle three spheres. The Bloch vector accumulates an angle with respect to the $\vec{z}$ axis, as shown by the upper middle three spheres, and is uninterrupted by collective spin flips, since these two rotations commute. Finally, an externally controlled magnetic gradient, rotates the Bloch vector about the $\vec{z}$ axis by $\phi_{parity}$ radians. The projection of the final Bloch vector on the $\vec{x}$ axis corresponds to the parity observable, measured by a collective $\pi/2$ rotation followed by projective state detection, as explained in the text.
    \textbf{e} Parity analysis fringe example (numerical).
    }
\end{figure}

\begin{figure} 
	\includegraphics[scale=0.95]{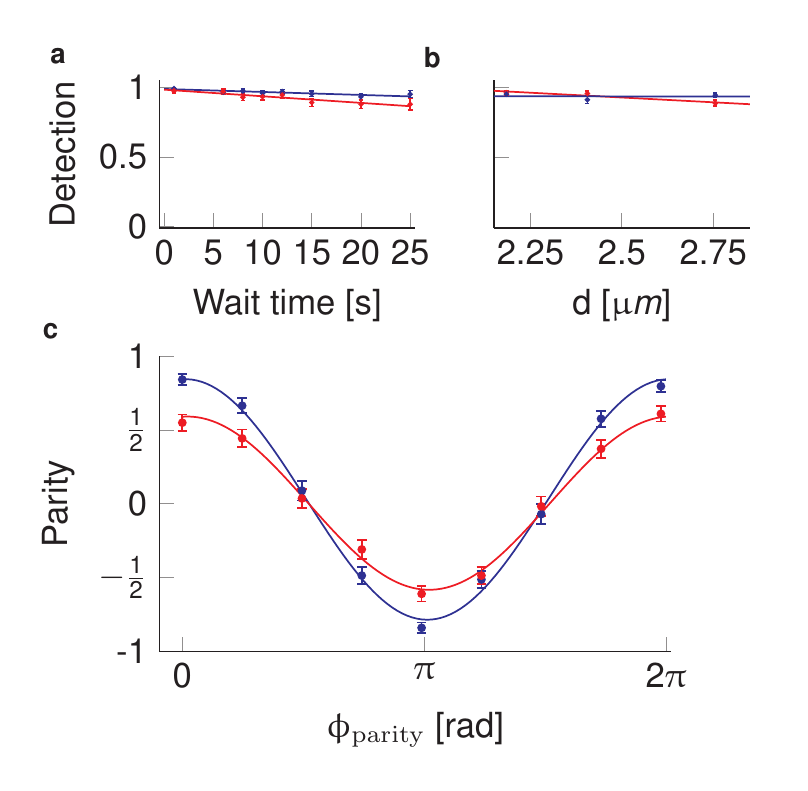}
	\caption{
	Characterization of quantum decoherence.
	\textbf{a}, Detection fidelity vs. experiment time at an inter ion distance of $d=2.4\ \mu m$. The probability of measuring $\kupup$ ($\kdowndown$) given that the system was initialized to $\kupup$ ($\kdowndown$) is shown by the red (blue) dots. Solid lines are linear best fits. During the experiment, collective spin flips are applied at a period of $0.5\ s$ as in the actual spin-spin experiment. Detection fidelities degrades due to ion motion heating.
    \textbf{b}, Detection fidelity vs. inter-electron distance $d$ at $T=15$s experiment (similar to a).
	\textbf{c}, Dephasing time estimate. The system is initialized to $\ket{\Psi_+}=(\kupdown+\downup)/\sqrt{2}$ followed by a train of spin echos as in \textbf{a} and \textbf{b}. Parity analysis is performed after a wait time of $T=1\ s$ ($15\ s$), shown by the blue (red) dots. Solid blue (red) line are a best fit to a cosine fringe, yielding an amplitude of $0.81(5)$ ($0.59(4)$). A conservative estimate for the dephsasing time, not taking detection degradation into account, yields $44\pm 12 s$. Taking into account the degradation in detection fidelity, as characterized in a, we observe no statistically significant dephasing.
	}
\end{figure}

\begin{figure} 
	\includegraphics{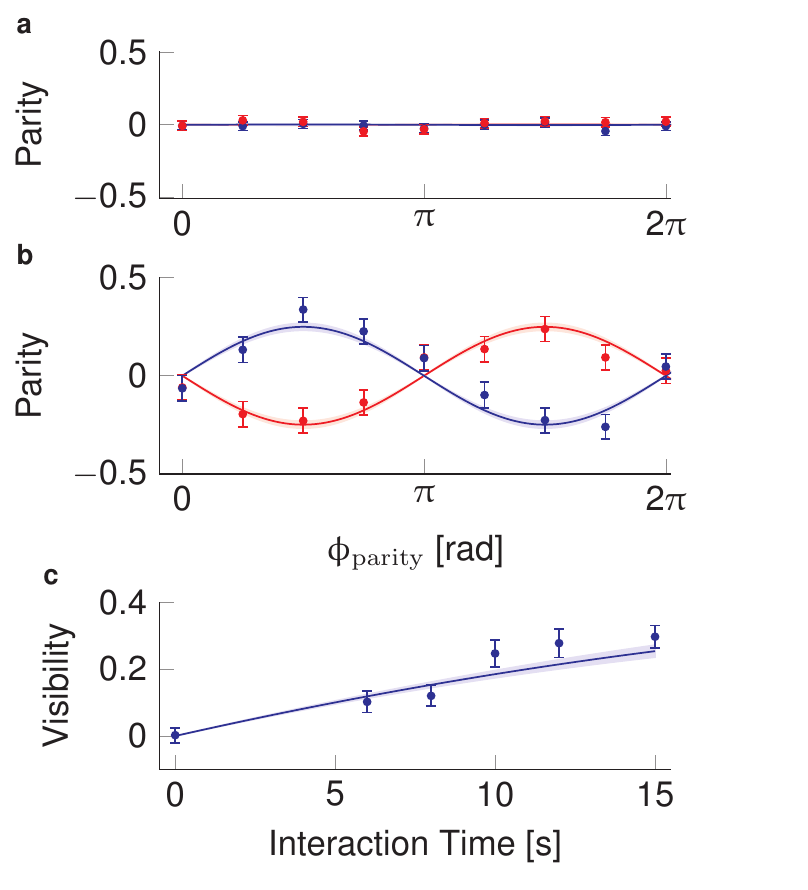}
	\caption{
	Coherent oscillations due to spin-spin interaction at an inter-electron distance of $d=2.4\ \mu m$.
	\textbf{a}, Parity analysis of a $0.1\ s$ long spin-spin experiment. Blue (red) dots show the parity measurements when the initial state is $\kupdown$ ($\kdownup$). Solid lines are the spin-spin theory with no adjustable parameters, taking into account the preparation and detection fidelities, characterized in Figure 2a.
	\textbf{b}, Same as Figure 3a, for a $T=15\ s$ long experiment. Shaded areas are one standard deviation intervals for the corresponding solid line spin-spin theory, resulting from preparation and detection probability estimate uncertainty. A best fit to $A\sin(\phi_{parity})$ (not shown) yields $0.24(3)$, from which the spin-spin coupling constant is estimated to be $\xi=(2\pi)0.9(1)$ mHz in reasonable agreement with theory ($\xi=(2\pi)0.93$ mHz).
	\textbf{c}, Parity amplitude (Visibility) vs. spin-spin interaction time $T$. The parity observable is measured at $\phi_{parity}=\pi/2$. Solid line and shaded area are the same as in \textbf{a,b}. A fit to $\alpha\sin((2\pi)4\xi T)$ (not shown) yields $\xi=1.1(2)\ mHz$. Here $\alpha$ is the visibility degradation factor, extracted from the data shown in Figure 2a, as explained in the text.
	}
\end{figure}

\begin{figure} 
	\includegraphics{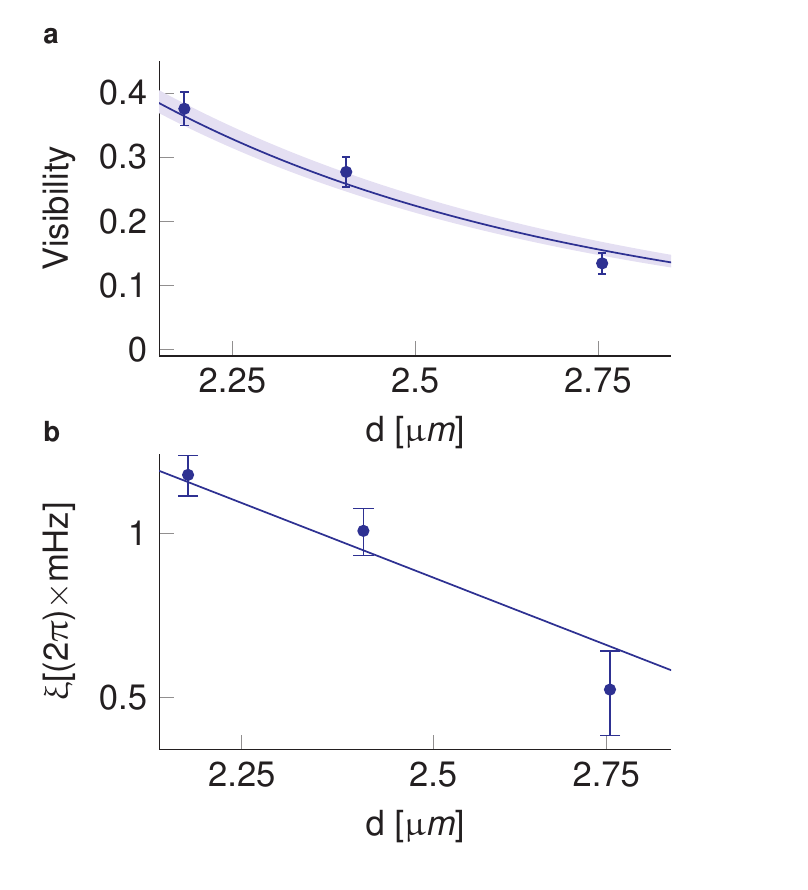}
	\caption{
	Spin-spin interaction as a function of the distance between the two ions.
    \textbf{a}, Parity visibility vs. ion separation $d$ is shown by the blue dots for a fixed experiment time $T=15$ s. Solid line is spin-spin theory without any adjustable parameters, taking preparation and detection fidelities into account, as characterized in Figure 2b. Shaded blue area is a one standard deviation interval for the solid line spin-spin theory, resulting from preparation and detection probability estimated uncertainty.
	\textbf{b}, Spin-spin coupling strength $\xi$ vs. ion separation (log-log scale). Blue dots are extracted from \textbf{a}, using $\Visibility=\alpha\sin(4\xi T)$. The Visibility degradation factor $\alpha$ is the extracted from the data in Figure 2b. Solid blue line is spin-spin theory without any adjustable parameters. A linear best fit to $\xi=\mu_0(g\mu_B/2)^2/4\pi \hbar d^n$ (not shown) yields $n=3.0(4)$, consistent with the $n=3$ theoretical exponent.
	}
\end{figure}

\end{document}